\documentclass[reprint, superscriptaddress, showpacs,amsmath, amssymb,aps,prl]{revtex4-2}
\usepackage{graphicx}
\usepackage{dcolumn}
\usepackage{bm}
\usepackage{upgreek}
\graphicspath{{figure/}{figsgaoerb/}}
\usepackage{hyperref}
\usepackage{booktabs, array, mathptmx, float, tabularx, booktabs, lipsum, amsmath,multirow}
\hypersetup{colorlinks=true, citecolor=blue, urlcolor=blue, linkcolor=blue}
\usepackage{color}

\begin{document}
\title{Improved Limits on an Exotic Spin- and Velocity-Dependent Interaction at the Micrometer Scale with an Ensemble-NV-Diamond Magnetometer}

\author{Diguang Wu}
\affiliation{CAS Key Laboratory of Microscale Magnetic Resonance and School of Physical Sciences, University of Science and Technology of China, Hefei 230026, China}
\affiliation{CAS Center for Excellence in Quantum Information and Quantum Physics, University of Science and Technology of China, Hefei 230026, China}
\affiliation{Hefei National Laboratory, University of Science and Technology of China, Hefei 230088, China}
\author{Hang Liang}
\affiliation{CAS Key Laboratory of Microscale Magnetic Resonance and School of Physical Sciences, University of Science and Technology of China, Hefei 230026, China}
\affiliation{CAS Center for Excellence in Quantum Information and Quantum Physics, University of Science and Technology of China, Hefei 230026, China}

\author{Man Jiao}
\email{jm2012@ustc.edu.cn}
\affiliation{CAS Key Laboratory of Microscale Magnetic Resonance and School of Physical Sciences, University of Science and Technology of China, Hefei 230026, China}
\affiliation{CAS Center for Excellence in Quantum Information and Quantum Physics, University of Science and Technology of China, Hefei 230026, China}

\author{Yi-Fu Cai}
\affiliation{CAS Key Laboratory for Researches in Galaxies and Cosmology, School of Astronomy and Space Science, University of Science and Technology of China, Hefei, Anhui 230026, China}
\affiliation{Department of Astronomy, School of Physical Sciences, University of Science and Technology of China, Hefei, Anhui 230026, China}

\author{ Chang-Kui Duan}
\affiliation{CAS Key Laboratory of Microscale Magnetic Resonance and School of Physical Sciences, University of Science and Technology of China, Hefei 230026, China}
\affiliation{CAS Center for Excellence in Quantum Information and Quantum Physics, University of Science and Technology of China, Hefei 230026, China}

\author{Ya Wang}
\affiliation{CAS Key Laboratory of Microscale Magnetic Resonance and School of Physical Sciences, University of Science and Technology of China, Hefei 230026, China}
\affiliation{CAS Center for Excellence in Quantum Information and Quantum Physics, University of Science and Technology of China, Hefei 230026, China}
\affiliation{Hefei National Laboratory, University of Science and Technology of China, Hefei 230088, China}

\author{Xing Rong}
\email{xrong@ustc.edu.cn}
\affiliation{CAS Key Laboratory of Microscale Magnetic Resonance and School of Physical Sciences, University of Science and Technology of China, Hefei 230026, China}
\affiliation{CAS Center for Excellence in Quantum Information and Quantum Physics, University of Science and Technology of China, Hefei 230026, China}
\affiliation{Hefei National Laboratory, University of Science and Technology of China, Hefei 230088, China}

\author{Jiangfeng Du}
\email{djf@ustc.edu.cn}
\affiliation{CAS Key Laboratory of Microscale Magnetic Resonance and School of Physical Sciences, University of Science and Technology of China, Hefei 230026, China}
\affiliation{CAS Center for Excellence in Quantum Information and Quantum Physics, University of Science and Technology of China, Hefei 230026, China}
\affiliation{Hefei National Laboratory, University of Science and Technology of China, Hefei 230088, China}

\date{\today}

\begin{abstract}
Searching for exotic interactions provides a path for exploring new particles beyond the standard model. Here, we used an ensemble-NV-diamond magnetometer to search for an exotic spin- and velocity-dependent interaction between polarized electron spins and unpolarized nucleons at the micrometer scale. A thin layer of nitrogen-vacancy electronic spin ensemble in diamond is utilized as both the solid-state spin quantum sensor and the polarized electron source, and a vibrating lead sphere serves as the moving unpolarized nucleon source. The exotic interaction is searched by detecting the possible effective magnetic field induced by the moving  unpolarized nucleon source using the ensemble-NV-diamond magnetometer. Our result establishes new bounds for the coupling parameter $f_\perp$ within the force range from 5 to 400 $\upmu$m. The upper limit of the coupling parameter at 100 $\upmu$m is $\lvert f_\perp \rvert \leq 1.1\times 10^{-11}$, which is 3 orders of magnitude more stringent than the previous constraint. This result shows that NV ensemble can be a promising platform to search for hypothetical particles beyond the standard model.
\end{abstract}
\maketitle

Exotic long-range spin-dependent interactions mediated by new bosons are related to potential solutions to various mysteries in fundamental physics \cite{safronova2018searcha}.  Axion is such a Nambu-Goldstone boson of the spontaneously broken Peccei-Quinn symmetry \cite{weinberg1978new,wilczek1978problem} and is initially introduced to solve the strong \emph{CP} problem in QCD \cite{peccei1977constraints,peccei1977mathrmcp,kim2010axions}. The existence of the axion or axionlike particles is a prominent explanation of several unsolved fundamental conundrums, such as the composition of the dark matter \cite{bertone2005particle}, the origin of the dark energy \cite{peebles2003cosmological, kamionkowski2014dark}, and the hierarchy problem \cite{graham2015cosmological}. Exotic long-range spin-dependent interactions mediated by axion were proposed to search for new physics \cite{moody1984new}. Subsequently, these exotic interactions were  extended to sixteen types of potentials involving the spin polarization and momenta of the interacting particles \cite{dobrescu2006spindependent}, and revisited in \cite{fadeev2019revisiting}. These exotic interactions can be  not only mediated by pseudoscalar bosons such as axions and axionlike particles but also mediated by axial-vector bosons such as dark photons and \emph{Z$'$} bosons \cite{dobrescu2006spindependent,leslie2014prospectsa}.

Recent progress of experimental searching for exotic interactions was advanced with a wide range of precision measurement techniques \cite{safronova2018searcha} utilizing atomic magnetometer \cite{kim2018experimental, kim2019experimental, ji2018newa, lee2018improveda, wu2022experimental}, spin-based amplifier \cite{su2021search, wang2022limits}, spectroscopic measurements \cite{ficek2017constraints, ficek2018constraints}, cantilever \cite{ding2020constraints}, and trapped ions \cite{kotler2015constraints}. These advanced precision measurements in the laboratory are powerful tools for tests of fundamental physics as complementary to astronomical observation. The single nitrogen-vacancy (NV) center in diamonds has emerged as a solid-state spin quantum sensor to search for exotic spin-dependent interactions at the micrometer scale \cite{rong2018constraints, rong2018searching, jiao2021experimental, rong2020observation}, taking advantage of the ability enabling close proximity between the sensor and the source. Recently, a nonzero result has been reported in an experimental searching for exotic interactions \cite{rong2020observation}. It is necessary to perform experiments with improved sensitivities to scrutinize whether the origin of the nonzero result is physical or instrumental.

\begin{figure*}[htbp]
\centering
\includegraphics[width=2\columnwidth]{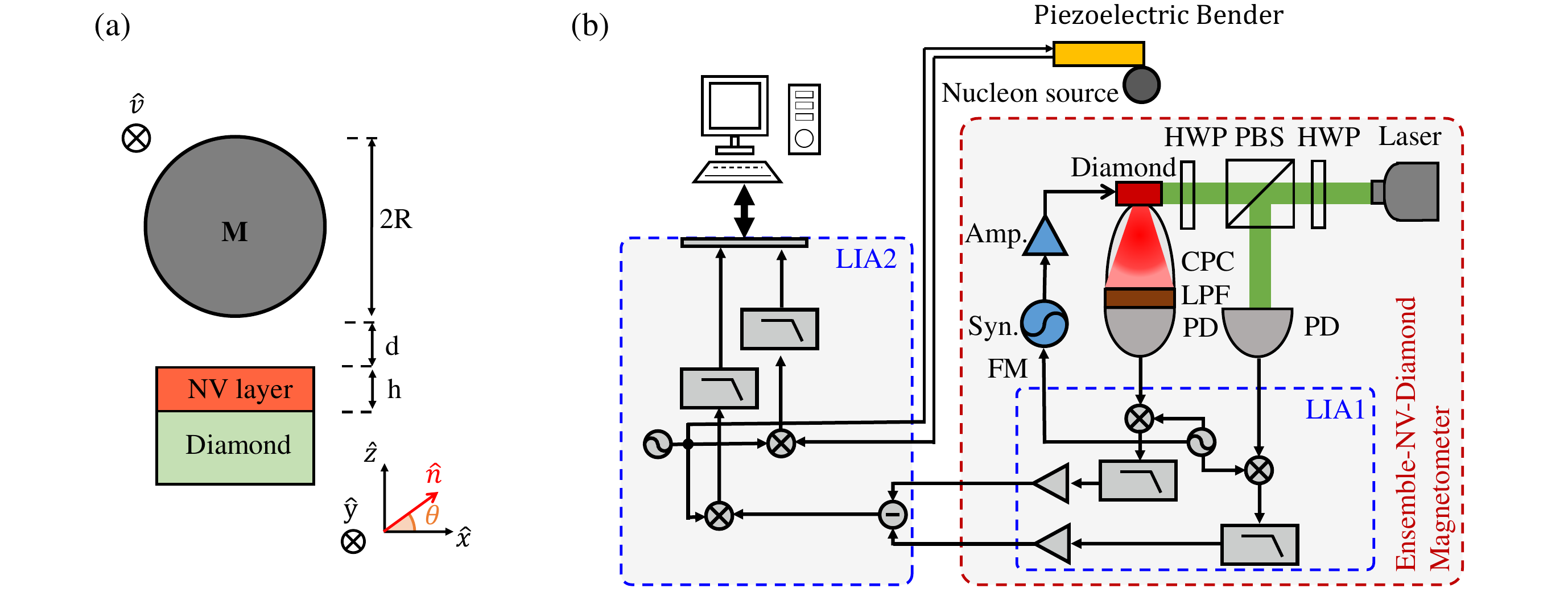}
    \caption{(a) Schematic experimental parameters. A lead sphere denoted as \emph{M} with the radius being \emph{R} vibrates parallel to the diamond surface. \emph{d} is the distance between the bottom of the sphere and the surface of the diamond. \emph{h} stands for the thickness of the NV layer. The direction of the NV axis is denoted as $\hat{n}$. $\hat{v}$ is the unit relative velocity vector between the lead sphere and the NV layer. (b) The scheme of the experimental setup based on the ensemble-NV-diamond magnetometer. NV centers are illuminated by a 532 nm laser which passes through two half-wave plates (HWP) and a polarizing beam splitter  (PBS).  The CPC is the compound parabolic concentrator, LPF is the long-pass filter, and PD is the photodiode. FM stands for the modulation frequency of the microwave from the synthesizer (Syn.). The output magnetic field measurement result of the magnetometer can be obtained by demodulating PD signal with the first lock-in amplifier (LIA1). The nucleon source is attached to a piezoelectric bender, which vibrates at the frequency $f_M$. The output of the magnetometer is further demodulated by the second lock-in amplifier (LIA2) to obtain the magnetic field component at the frequency $f_M$. The operations in each commercial LIA are displayed schematically in the figure.}
    \label{SAsetup}
\end{figure*}

In this work, we develop a technique based on an ensemble-NV-diamond magnetometer to search for an exotic spin- and velocity-dependent interaction between polarized electron spins and unpolarized nucleons. An ensemble of NV centers in diamond serves as both polarized electron sources and sensitive sensors.  The extension from single NV centers to NV ensemble provides better magnetic detection sensitivity, which makes the ensemble-NV-diamond magnetometer a promising platform for the searches of exotic interactions \cite{barry2020sensitivitya, liang2022experimental}.  Our technique also takes the advantage that, by averaging the magnetic field sensed by each NV center, the NV ensemble is insensitive to the effect of the diamagnetism of the material of the nucleon source \cite{liang2022experimental}.
Here, we focus on the exotic spin- and velocity-dependent interaction between an electron spin and a nucleon ($V_{4,5}$ in Ref. \cite{dobrescu2006spindependent}’s notation), described as
\begin{equation}
V_{4,5}=-f_\perp\frac{\hbar^2}{8\uppi m_e\mathrm{c}}{\hat{\sigma}\cdot}( \vec{v}\times \hat{r})\left(\frac{1}{\lambda r}+\frac{1}{r^2}\right)e^{-r/\lambda},
\end{equation}
where  $f_\perp$ is the dimensionless interaction coupling  parameter, $\hbar$ is the reduced Planck's constant, $m_e$ is the mass of the electron, $c$ is the speed of light in vacuum, $\hat{\sigma}$ is the Pauli vector of the electron spin. $\vec{v}$ is the relative velocity vector, and $\lambda=\hbar/(m_bc)$ is the force range with $m_b$ being the mass of the hypothetical boson. $\vec{r}$ is the displacement vector between the electron and the nucleon, $r=|\vec{r}|$ and $\hat{r}={\vec{r}}/{r}$. The exotic spin- and velocity-dependent interaction can be mediated by the axion or the \emph{Z$'$} boson \cite{leslie2014prospectsa,dobrescu2006spindependent}. The effective Lagrangian is $\mathcal{L}_a=-a\sum_\psi \bar{\psi}\gamma^\mu(g_S^\psi+\gamma_5g_P^\psi)\psi$ for axion, or
$\mathcal{L}_{Z'}=Z'_\mu\sum_\psi \bar{\psi}\gamma^\mu(g_V^\psi+\gamma_5g_A^\psi)\psi$ for \emph{Z$'$} boson, where $\psi$ stands for the fermion field, $\gamma^\mu$ and $\gamma_5$ are Dirac matrices. For axion exchange,  $f_\perp=g_s^eg_s^N/2$, where $g_s^e\ (g_s^N)$ denotes the scalar electron (nucleon) coupling. In the case of \emph{Z$'$} boson, $f_\perp=g_v^eg_v^N/2$, where  $g_v^e\ (g_v^N)$ is the  vector electron (nucleon) coupling \cite{leslie2014prospectsa,dobrescu2006spindependent}. The exotic interaction induces an effective magnetic field sensed by the electron spin
\begin{equation}
\begin{aligned}
B_{\mathrm{eff}} (r)=-f_\perp&\frac{\hbar}{4\uppi m_e c \gamma_e}\hat{n}\cdot(\vec{v}\times\hat{r})\left(\frac{1}{\lambda r}+\frac{1}{r^2}\right)e^{-r/\lambda} ,
\end{aligned}
\end{equation}
where $\gamma_e=2\uppi\times28$ GHz/T is the gyromagnetic ratio of the NV electron spin.

Figure \ref{SAsetup}(a) shows the geometric parameters of  the spin sensor and the nucleon source. A 23-$\upmu$m-thick layer of NV ensemble at the surface of the diamond is used as the spin sensor. The size of the layer is $660\times661\times23\ \upmu$m$^3$. The substrate is a single crystal $\langle100\rangle$ orientated diamond with electronic grade high purity and parts-per-billion nitrogen density. The angle between the NV axis and the surface of the diamond is $\theta=\arcsin(1/\sqrt{3})$. A lead sphere with the diameter being $2R=897(3)\ \upmu$m is used as the unpolarized nucleon source. The nucleon density is $6.8\times10^{30}\ \rm m^{-3}$ \cite{lee2018improveda}. The lead sphere is attached to the end of a piezoelectric bender, which can vibrate in the direction parallel to the diamond surface. The NV axis is perpendicular to the velocity vector of the lead sphere. The amplitude of the vibration is $A=538(3)$ nm and the frequency is $f_M=941$ Hz. The distance between the bottom of the sphere and the surface of the diamond is $d=5.0(5)\ \upmu$m. The experimental setup based on an ensemble NV magnetometer is shown in Fig.~\ref{SAsetup}(b). The NV ensemble is polarized by a 532-nm laser with a diameter being 0.8 mm. The fluorescence from the NV ensemble is collected by a compound parabolic concentrator and converted to the photocurrent by a photodetector  (PD). A long-pass filter  (LPF) is used to filter out the laser from the fluorescence. A polarizing beam splitter (PBS) and another PD are used to monitor the power fluctuation of the laser. The laser polarization direction and the split ratio of PBS are adjusted by the two half-wave plates (HWP).

\begin{figure}[htbp]
\centering
\includegraphics[width=1\columnwidth]{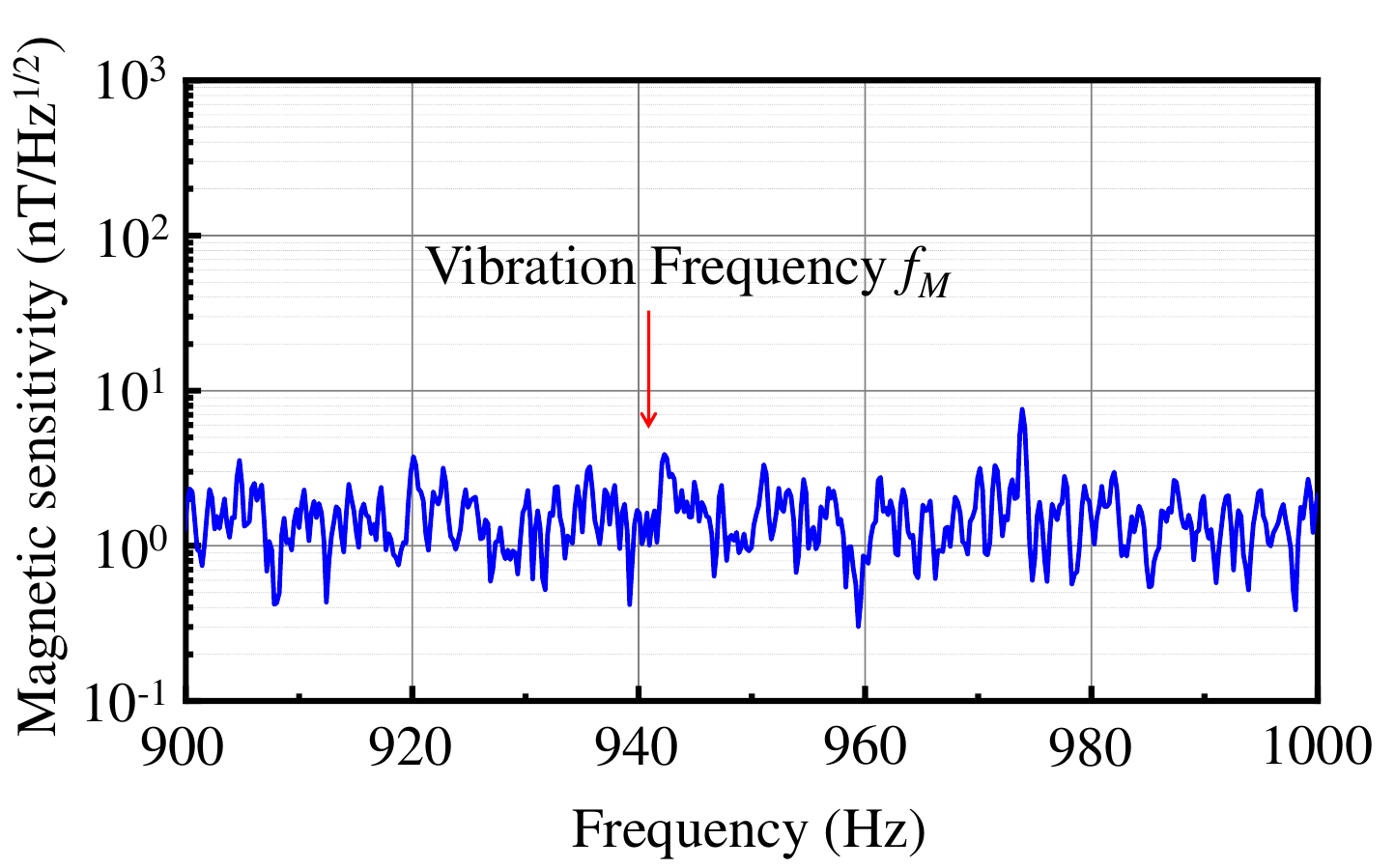}
    \caption{The magnetic sensitivity of the ensemble-NV-diamond magnetometer. The magnetic sensitivity is $1.6\ \rm nT/\sqrt{Hz}$ within the frequency range from 0.9 to 1 kHz. The vibration frequency $f_M=941$ Hz is marked with a red arrow.}
    \label{SAcali}
\end{figure}

The NV center is featured by a ground electronic spin triplet state with substates $|m_s=0\rangle$ and $|m_s=\pm1\rangle$ \cite{doherty2013nitrogenvacancy}. The  optical spin initialization into $|m_s=0\rangle$ and spin-state-dependent fluorescence contrast lead to the optically detected magnetic resonance \cite{goldman2015stateselective}. An ensemble-NV-diamond magnetometer based on the continuous-wave (CW) method \cite{liang2022experimental, xie2021hybrid, barry2016optical} is used to search for the exotic interaction. The optical polarization, the microwave excitation, and the fluorescence readout occur simultaneously in this CW magnetometry. A static bias magnetic field along the axis of NV centers is applied to remove the degeneracy of $|m_s=\pm1\rangle$. The fluorescence of the NV ensemble varies when a change in the local magnetic field shifts the resonance frequency. To avoid the flicker noise, the frequency of the microwave is modulated with modulation frequency of 87.975~kHz. The magnetic field sensed by the NV ensemble is encoded in the amplitude of the fluorescence intensity at the modulation frequency. The signals from PDs are demodulated by the first lock-in amplifier  [LIA1 in Fig.\ref{SAsetup}(b)] with the time constant being 8 $\upmu$s. Since the effects from laser power instability are common for signals from both PDs, these effects are suppressed by differential measurement \cite{schloss2018simultaneousa}. The magnetic sensitivity of the ensemble-NV-diamond magnetometer is shown in Fig.~\ref{SAcali}. The coefficient between the magnetometer output voltage signal and the magnetic field is $\eta=47.2\pm0.2\ \upmu$T/V. The magnetic sensitivity is $1.6\ \rm nT/\sqrt{Hz}$ within the frequency range from 0.9 to 1 kHz   (see Appendix \ref{SM2} for details).

The effective magnetic field $b_{\mathrm{eff}}$ sensed by the NV ensemble can be derived from integrating $B_{\mathrm{eff}} (r)$ over all nucleons and electron spins \cite{liang2022experimental}.
Since the nucleon source vibrates at a fixed frequency $f_M$, the velocity of the sphere can be represented as $v(t)=2\pi f_M A \sin{(2\pi f_Mt)}$, where \emph{A} is the amplitude of vibration. The Fourier transform of the effective magnetic field is
\begin{equation}
\begin{aligned}
b_{\mathrm{eff}}(t)=\sum_{n=1}^\infty b_{n}\sin(2\pi nf_Mt).
\end{aligned}
\end{equation}
where $b_n$ is the amplitude of the $n$th harmonic. Here, we focus on the measurement of the amplitude of the first-order harmonic of the velocity-dependent effective magnetic field $b_1$. The output signal of LIA1 is demodulated by a second lock-in amplifier [LIA2 in Fig.\ref{SAsetup}(b)]. The experimentally calibrated phase delay between the output of magnetometer and the vibration of nucleon source at $f_M$ is $\phi=51\pm4^\circ$. The method of the phase calibration is the same as Ref. \cite{liang2022experimental}. The time constant of LIA2 is 71~ms. The quadrature component of the demodulated signal from LIA2 corresponds to $b_1$ (see Appendix \ref{SM3} for details).

\begin{figure}[htbp]
\centering
\includegraphics[width=0.9\columnwidth]{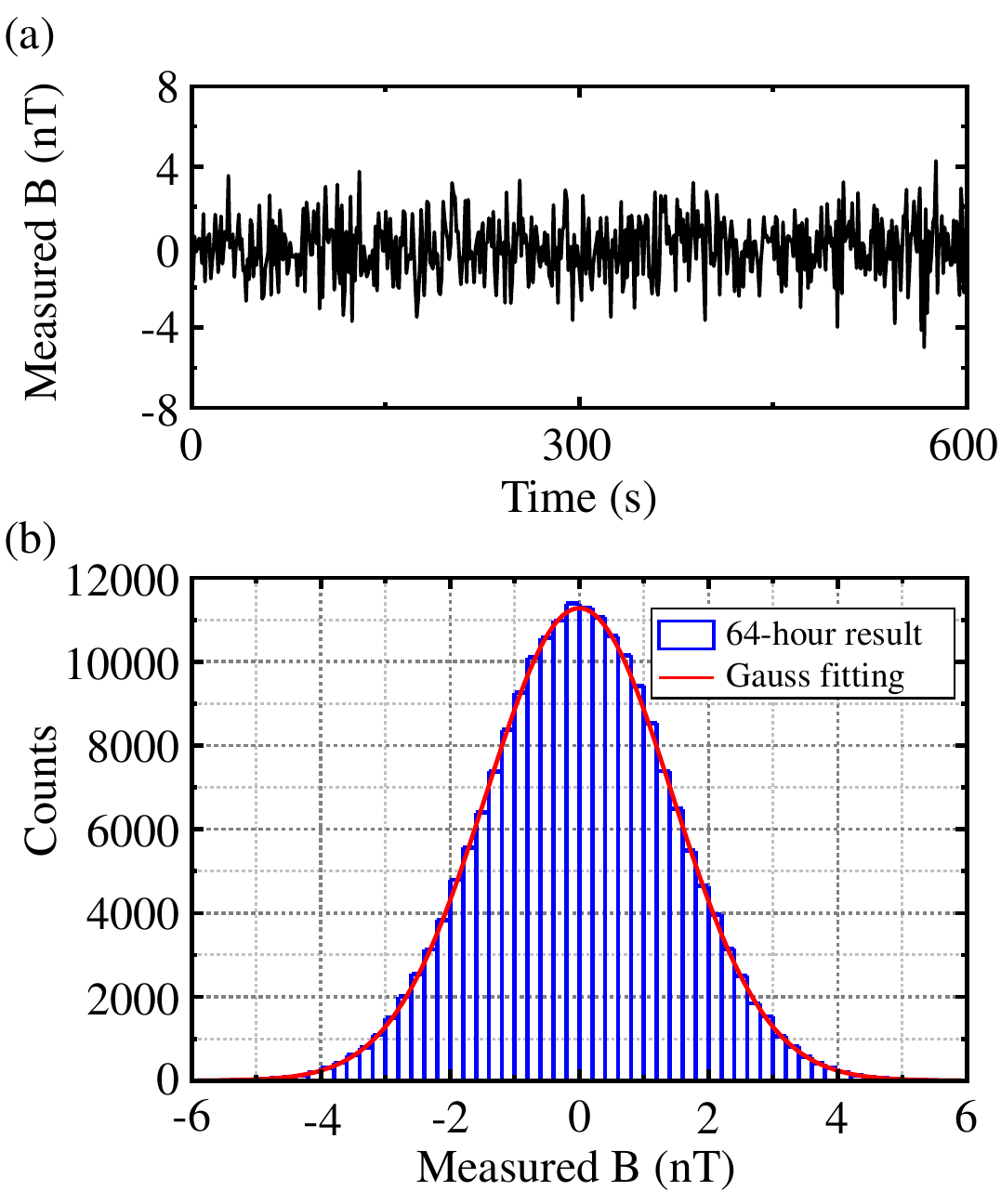}
    \caption{(a) The black line shows the time trace of the experimentally measured effective magnetic fields for 600 s. (b) The histogram of the experimentally measured effective magnetic fields for 64-h collection. The red line indicates the fit to a Gaussian distribution. The total counts are 0.2 M, the mean of the result is $1.0$ pT, and the standard error of the mean is $3.2$ pT.}
    \label{SAresult}
\end{figure}

A typical time trace of the experimentally measured quadrature part of the output of LIA2 is shown in Fig.~\ref{SAresult}(a) with the time duration being 600 s. The measurement was carried out for 64 h to reduce the statistical uncertainty. A histogram of the measured effective magnetic fields collected for 64 h is shown in Fig.~\ref{SAresult}(b). The fit shows that the result obeys the Gaussian distribution. The measured first-order harmonic component of the effective magnetic fields is $b_1^\mathrm{exp}=1.0\pm 3.2$ pT and the corresponding coupling  parameter $f_\perp$ at $\lambda=100\ \upmu$m is $(1.5\pm4.9)\times10^{-12}$. The result that the mean value of the measured magnetic field is less than the standard error shows no evidence of the exotic interaction in this experiment. Therefore, new limits of the coupling  parameter  $f_\perp$ can be obtained.

Table~\ref{t3} summarizes the systematic errors of our experiment. The uncertainties of experimental parameters and the corrections to the coupling  parameter $f_\perp$ with $\lambda=100\ \upmu$m are provided. The main sources of the systematic errors include the uncertainties of the diameter of the sphere $2R$, the thickness of the NV layer $h$, the vibration amplitude $A$, the angle $\alpha$ between the direction of the velocity and the $\hat{y}$ direction, the distance $d$ between the bottom of the sphere and the surface of the diamond, the phase of the reference signal $\phi$, nucleon density, the misalignment between the center of the sphere and the center of NV centers in the \emph{x-y} plane, and the coefficient $\eta$ between the magnetometer output voltage signal and the magnetic field (see Appendix \ref{SM5} for details). By quadrature summing the items of all systematic errors, the total systematic error for the coupling  parameter is $\pm2\times 10^{-14}$ with $\lambda=100\ \upmu$m. Taking both statistical and systematic errors into account, the bound for the coupling  parameter with $\lambda=100\ \upmu$m is $\lvert f_\perp \rvert\leq 1.1\times 10^{-11}$ with a 95\% confidence level. The other value of the bound with the different force range can be obtained with the same method.

\begin{table}[h!]
 \caption{Summary of systematic errors. The corrections to  $f_\perp$ with $\lambda=100\ \upmu$m are listed.}
 \label{t3}
\begin{tabular}{l c c}
\hline
\hline
Parameter&Value&$\Delta f_\perp (10^{-13})$\\
\hline
Diameter 2R & $ 897\pm 3\ \upmu$m&$\pm0.05$ \\
Thickness $h$& $23 \pm 1 \upmu$m &$\pm0.08$ \\
Amplitude $A$& $538\pm 3$ nm & $\pm0.09$\\
Angle $\alpha$ &$0 \pm3 ^\circ$ &$^{+0.02}_{-0.00}$ \\
Distance $d$& $5.0\pm 0.5\ \upmu$m & $\pm0.08$\\
Phase $\phi$& $51\pm 4^\circ$ & $_{-0.00}^{+0.03}$\\
Nucleon density&$(682\pm2)\times10^{28}\ \mathrm{m^{-3}}$ & $\pm0.05$\\
Centroid deviation&\multirow{2}{1.7 cm}{\centering$0\pm 10\ \upmu$m} &\multirow{2}{1.7 cm}{\centering$^{+0.09}_{-0.08}$} \\
$\ $ in \emph{x-y} plane& & \\
Coefficient $\eta$ &$47.2\pm0.2\ \upmu$T/V&$\pm0.07$ \\
\hline
\multirow{2}{3 cm}{Obtained $f_\perp$} &\multirow{2}{1.7 cm}{$15.4\times10^{-13}$}&$\pm 49.3$ (statistic) \\
 & &$\pm 0.2$ (systematic)\\
\hline
\hline
\end{tabular}
\end{table}

\begin{figure}[htbp]
\centering
\includegraphics[width=1\columnwidth]{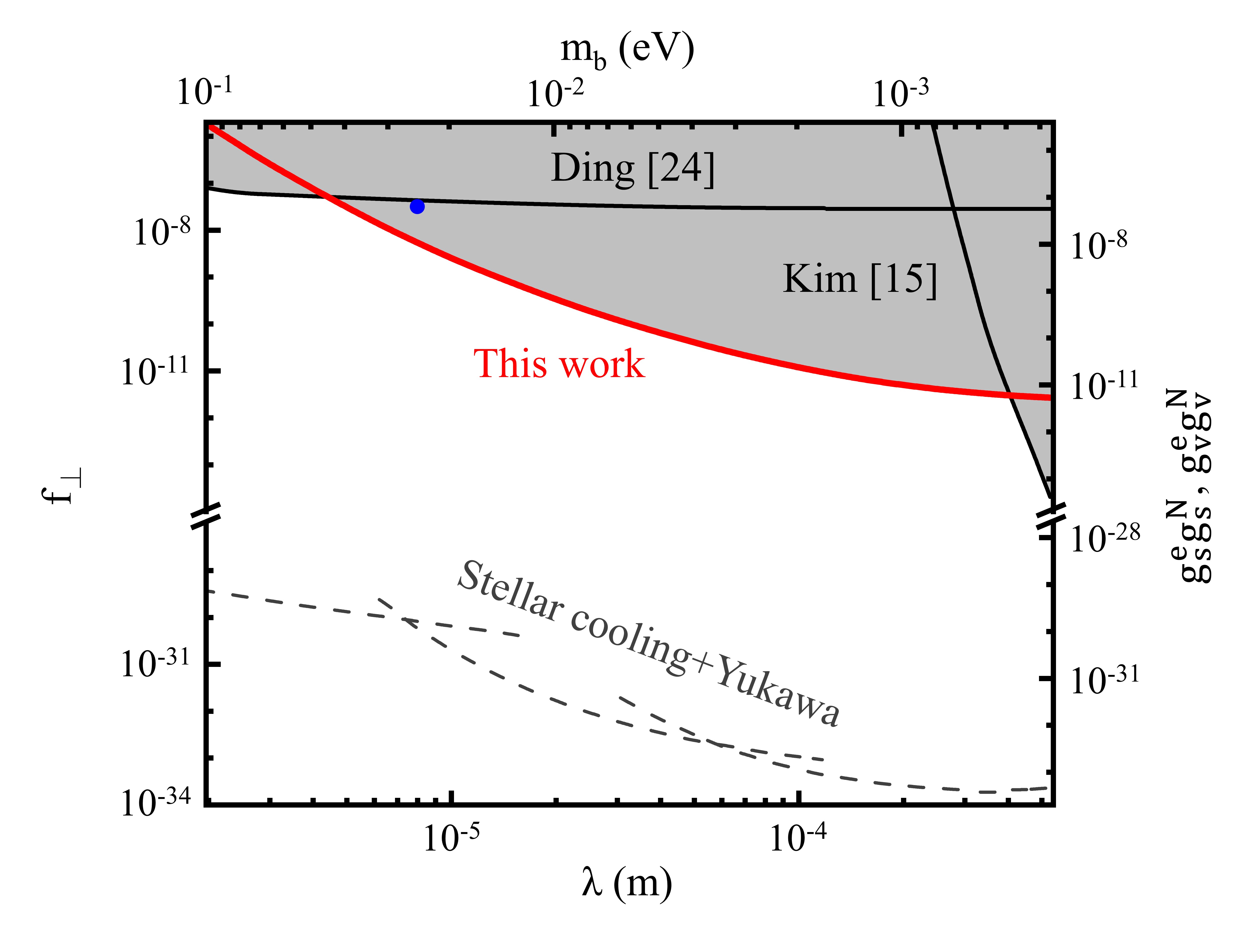}
    \caption{The upper bounds of the coupling  parameter $f_\perp$ as a function of the force range $\lambda$ and the boson mass $m_b$. Black lines are upper bounds obtained from previous experiments \cite{kim2018experimental,ding2020constraints}. The blue dot denotes the nonzero result reported in Ref.\cite{rong2020observation}. The red line is the upper limit established by our experiment. The gray-filled area shows the excluded values. More stringent experimental bounds are obtained by our experiment within the force range from 5 to $400\ \upmu$m. The dashed line is set by a  combination of stellar cooling rate observation and tests of the hypothetical Yukawa interactions \cite{raffelt2012limits,chen2016stronger,tan2020improvement,lee2020new}.}
    \label{SAbound}
\end{figure}

Figure \ref{SAbound} shows the upper bounds of the coupling  parameter  $f_\perp$ of the exotic spin- and velocity-dependent interaction established by this work together with previous experimental constraints \cite{kim2018experimental, ding2020constraints}. For the force range $\lambda <5\ \upmu$m, the limits of  the coupling   parameter $f_\perp$  were obtained by Ding \textit{et al.} with a cantilever to detect the force between a vibrating gold sphere and a microfabricated magnetic structure \cite{ding2020constraints}. The constraints for the force range $\lambda>400\ \upmu$m were set by Kim \textit{et al.} with a spin-exchange-relaxation-free magnetometer to detect the possible effective magnetic field induced by a rotating BGO crystal \cite{kim2018experimental}. For the force range from 5 to $400\ \upmu$m,  improved experimental bounds are established by our experiment. The upper bound for the force range $\lambda=100\ \upmu$m is  $\lvert f_\perp \rvert\leq1.1\times 10^{-11} $, which is more than 3 orders of magnitude more stringent than the previous laboratory constraint \cite{ding2020constraints}. The constraints of stellar cooling are derived by a combination of $g_s^e$ from the stellar cooling \cite{raffelt2012limits} and $g_s^N$  from experimental tests of the hypothetical Yukawa interactions \cite{chen2016stronger,tan2020improvement,lee2020new}.  Notice that astrophysical bounds on $g_s^e$ may suffer from the uncertainties, such as the accuracy of stellar models \cite{hardy2017stellar}, degree of model specificity \cite{masso2005evading}, and invalidation due to chameleon mechanism \cite{jain2006evading}.

In summary, we  conducted an experimental search for an exotic spin- and velocity-dependent interaction between polarized electron spins and unpolarized nucleons  with an ensemble-NV-diamond magnetometer. Improved constraints of the coupling  parameter  $f_\perp$ have been established within the force range from 5 to $400\ \upmu$m. This work is also an experimental test of the nonzero result in Ref. \cite{rong2020observation}. Different from the single NV setup in Ref. \cite{rong2020observation}, by averaging the magnetic field sensed by each NV center, our setup is insensitive to the effect due to the diamagnetism of the nucleon source \cite{liang2022experimental}. We carefully synchronized the ensemble-NV-diamond magnetometer and the vibration of the nucleon source using a calibration signal as in Ref. \cite{liang2022experimental}. This setup is more sensitive than the single NV magnetometer in Ref. \cite{rong2020observation} with the force range larger than 8 $\upmu$m.  Our result suggests that the nonzero result in Ref. \cite{rong2020observation} may come from instrumental artifacts (seeAppendix \ref{SM6} for details).  In the future, more stringent constraints can be obtained by further improving the sensitivity of the ensemble-NV-diamond magnetometer. The dynamical decoupling technology can be utilized to optimize the dephasing time \cite{zhang2021diamond, delange2010universala}. The readout fidelity can be improved by the near-infrared absorption readout \cite{chatzidrosos2017miniature}.  Besides, coating the diamond surface is beneficial to increase the fluorescence collection efficiency \cite{yu2020enhanceda}.  The nitrogen delta-doping technique can be used to fabricate the shallow NV centers with a nanoscale thickness \cite{ohno2012engineering}, enabling searching for the exotic interactions with a shorter force range. Besides, the method based on an ensemble-NV-diamond magnetometer can be utilized to explore other types of exotic interactions with additional functional upgradation in the future. Furthermore, high-density NV ensembles can be used not only as spin sensors but also as spin sources with well-controlled spin polarization. This work shows that the NV ensemble is a promising platform for searching for exotic spin-dependent interactions beyond the standard model.

This work was supported by the Chinese Academy of Sciences (Grants No. XDC07000000, No. GJJSTD20200001, No. QYZDY-SSW-SLH004, No. QYZDB-SSWSLH005), the National Key R\&D Program of China (Grant No. 2018YFA0306600, No. 2021YFC2203100), Anhui Initiative in Quantum Information Technologies (Grant No. AHY050000), Innovation Program for Quantum Science and Technology (2021ZD0302200), NSFC (11961131007, 11653002, 12150010,12205290,12261160569), China Postdoctoral Science Foundation (2022TQ0330), and the Hefei Comprehensive National Science Center. X. R. thanks the Youth Innovation Promotion Association of Chinese Academy of Sciences for the support. Y. F. C., Y. W., and M. J. thank the Fundamental Research Funds for Central Universities. Y. F. C. is supported in part by the CSC Innovation Talent Funds, by the USTC Fellowship for International Cooperation, and by the USTC Research Funds of the Double First-Class Initiative. This work was partially carried out at the USTC Center for Micro and Nanoscale Research and Fabrication.

D.G.W. and H.L. contributed equally to this work.

\appendix

\section{Devices in this experimental setup}
\label{SM1}
\begin{figure}[htbp]
\centering
\includegraphics[width=0.9\columnwidth]{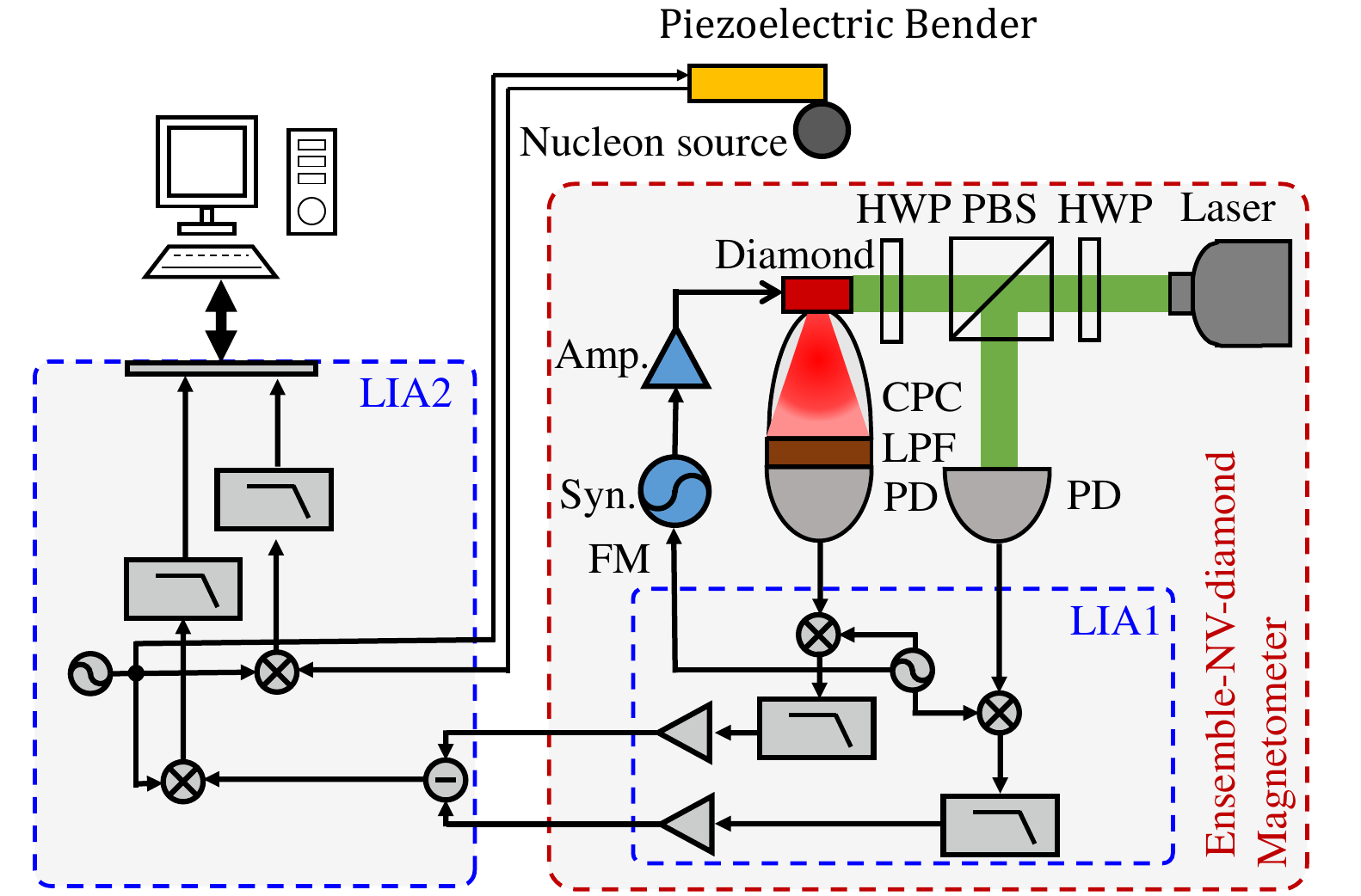}
    \caption{The scheme of the experimental setup based on the Ensemble-NV-Diamond magnetometer. The magnetometer is composed of diamond, half-wave plates (HWP), polarizing beam splitter (PBS), compound parabolic concentrator (CPC), the long-pass filter (LPF),  photodiode (PD), synthesizer(Syn.), amplifier (amp.), and the first commercial lock-in amplifier (LIA1). A piezoelectric bender drived by the second commercial lock-in amplifier (LIA2) is used to create the relative velocity term in the exotic velocity-dependent interaction. LIA2 is used to extract the magnitude of the first-order harmonic component of the effective magnetic field. The operations in each commercial LIA are displayed schematically in the figure.}
    \label{SMsetup}
\end{figure}

Figure~\ref{SMsetup} shows the scheme of the experimental setup based on an Ensemble-NV-Diamond Magnetometer, which is the same as Fig.1(b) in the main text. The manufacturers and models of devices used in this setup are shown in Table~\ref{devices}.
\begin{table}[h!]
 \caption{List of the detailed information about the devices in our experiment}
 \label{devices}
\begin{tabular}{l c l c l}
\hline
\hline
\textbf{Instrument}&\textbf{Manufacturer}&\textbf{Model}\\
\hline
Lock-in Amplifier 1 and 2& Zurich Instruments&HF2LI\\
Laser &Cobolt& 0532-05-01-1500-700\\
Synthesizer &National Instruments &FSW-0010\\
MW Amplifier&CIQTEK&GYPA2530-42\\
PD& Thorlabs& SM05PD1A\\
Piezoelectric Bender &Thorlabs & PB4VB2W\\
\hline
\hline
\end{tabular}
\end{table}

\section{The coefficient $\eta$ and The sensitivity of the ensemble-NV-diamond magnetometer}
\label{SM2}
\begin{figure}[htbp]
\centering
\includegraphics[width=1\columnwidth]{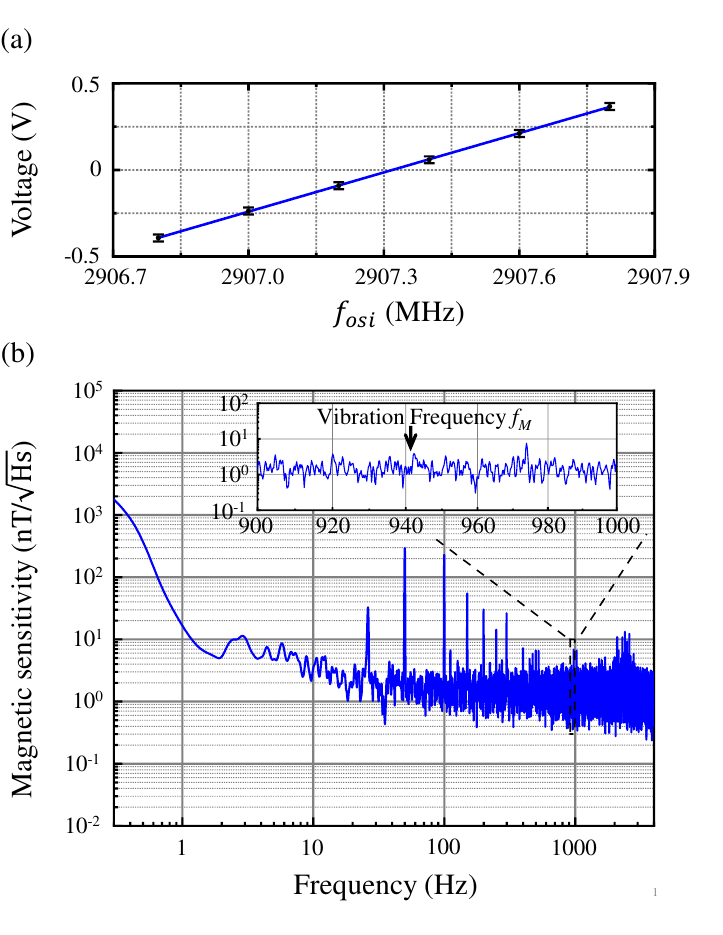}
    \caption{(a) The solid black circle points indicate the voltage from LIA1 with different microwave frequencies. The blue line is the linear fitting. The slope is $0.7563\pm0.0036$ V/MHz. (b) The magnetic sensitivity of the magnetometer. The power supply causes the peaks at 50 Hz and harmonics. The zoomed-in view shows the vibration frequency $f_M=941$ Hz.}
    \label{SMperform}
\end{figure}

The time-domain signal in voltage from the ensemble-NV-diamond magnetometer  is converted to the magnetic field by the coefficient $\eta$ between the magnetometer output voltage signal and the magnetic field. The specific region of the CW spectrum is shown in Fig.~\ref{SMperform}(a). The coefficient $\eta$ is $47.2\pm0.2\ \upmu$T/V according to the max slope of the CW spectrum $|\frac{\partial S}{\partial( f_{osi}-\gamma_eB)}|=0.7563\pm0.0036$ V/MHz and the gyromagnetic ratio of the NV center $\gamma_e=2\uppi\times28$ GHz/T. The sensitivity of our setup is obtained by the amplitude spectral density of the time-domain signal  \cite{xie2021hybrid}. Fig.~\ref{SMperform}(b) shows the magnetice detection sensitivity of the magnetometer. The sensitivity is $1.6\ \rm nT/\sqrt{Hz}$ within the frequency range from 0.9 to 1 kHz. Furthermore, the magnetic fields with unstable phases don't contribute after averaging over a long time. Because of the above reasons, the contribution of the static external magnetic field to the magnetic field at the vibration frequency is neglectable. Moreover, there is no significant magnetic field with the vibration frequency in phase with the velocity or the position.

\section{Separating velocity-dependent magnetic signals from position-dependent magnetic signals}
\label{SM3}

The effective magnetic field sensed by the NV ensemble is given by:
\begin{equation}
\begin{aligned}
b_{\mathrm{eff}}=-f_\perp \frac{\hbar \rho}{4\uppi m_e c \gamma_eV_S}\int_{S}\mathrm{d}V_S\int_{N} \hat{n}\cdot(\vec{v}\times\hat{r})(\frac{1}{\lambda r}+\frac{1}{r^2})e^{-\frac{r}{\lambda}}\mathrm{d}V_N
\end{aligned}
\end{equation}
where $V_S$ stands for the volume of the NV ensemble, $V_N$ is the volume of the nucleon source, S (N) indicates the volume integral over the space of the spin ensemble (nucleon source), $\vec{r}$ is the displacement vector between the electron and the nucleon, $r=|\vec{r}|$ and $\hat{r}={\vec{r}}/{r}$. $\gamma_e=2\uppi\times28$ GHz/T is the gyromagnetic ratio of the NV center, $\rho$ is the nucleon density, and $\hat{n}$ is the unit vector along the NV axis. The position of  the center of the lead sphere relative to the surface of diamond is $\vec{r}_M=A\cos(2\pi f_Mt)\hat{y}+(d+R)\hat{z}$. The velocity of sphere is $\vec{v}(t)=-2\pi f_MA\sin(2\pi f_M t)\hat{y}$.

Obviously, $b_{\mathrm{eff}}(t)=b_{\mathrm{eff}}(t+1/f_M)$ because the sphere is at the same position with the same velocity. It is noticed that $b_{\mathrm{eff}}(t)=-b_{\mathrm{eff}}(-t)$, because the sphere is at the same position with the opposite velocity. The Fourier transform of $b_{\mathrm{eff}}(t)$ is $b_{\mathrm{eff}}(t)=\sum_{n=1}^\infty b_{n}\sin(2\pi nf_Mt)$.

If a magnetic field $b_{\mathrm{other}}(t)$ arises from magnetic nanoparticles,  the magnetism of sphere, or energized coil, the magnetic field NV sensed is dependent on the relative position and is independent on the velocity. $b_{\mathrm{other}}(t)=b_{\mathrm{other}}(t+1/f_M)$ and $b_{\mathrm{other}}(t)=b_{\mathrm{other}}(-t)$ because the positions are the same. The result of Fourier expansion of $b_{\mathrm{other}}(t)$ is $b_{\mathrm{other}}(t)=\sum_{n=0}^\infty a_{n}cos(2\pi nf_Mt)$.

Therefore, the fundamental frequency component of $b_{\mathrm{eff}}$ is in-phase with the velocity $v(t)$ and the fundamental frequency component of $b_{\mathrm{other}}$ is in-phase with the position $r_M(t)$.

The signal from LIA1 is demodulated by LIA2. By adjusting the phase of internal reference of LIA2 to match the phase delay between the output of magnetometer and the vibration $\phi$, the amplitude of  the fundamental frequency component of $b_{\mathrm{eff}}$ relates to output from the quadrature channel of LIA2 and the amplitude of the fundamental frequency component of $b_{\mathrm{other}}$ relates to output from the  in-phase channel of LIA2. It suppresses the influence of the magnetic field which is dependent on the relative position and independent on the velocity.

\begin{figure}[htbp]
\centering
\includegraphics[width=0.9\columnwidth]{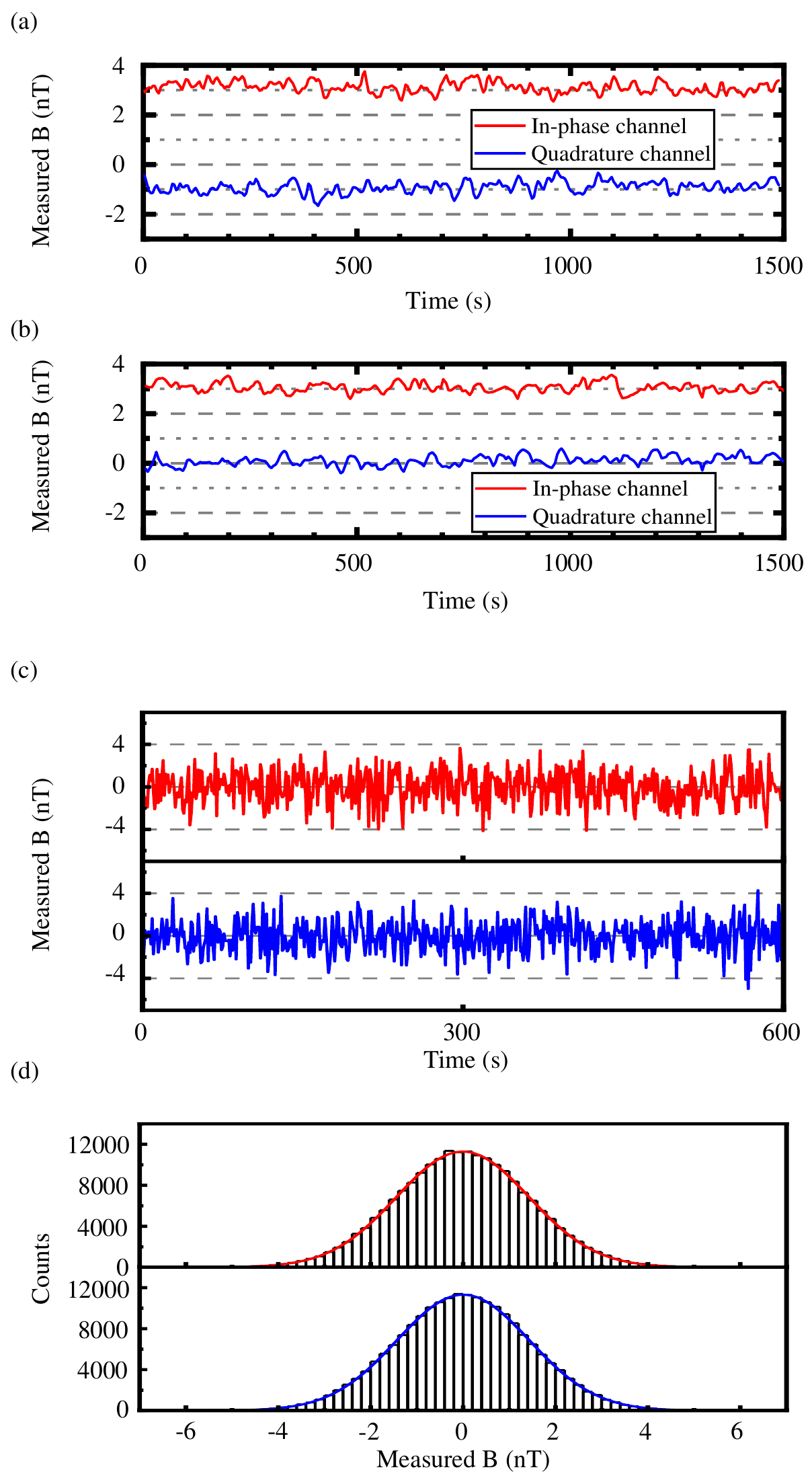}
    \caption{(a)(b) The time trace of  two components when a current is applied to generate a magnetic field that is strong enough to cover the effective magnetic field due to the exotic interaction. The red line indicates the in-phase component and the blue line indicates the quadrature component. (a) Before adjusting $\phi$, the signal due to current is output through both the in-phase channel and the quadrature channel. (b) After adjusting $\phi$,  the signal due to current is output through the in-phase channel.  (c)The result of the measured magnetic field with the time duration being 600 s. The upper one is output through the in-phase channel and the bottom one is output through the quadrature channel. (d) The Histogram of the measured magnetic fields which are collected for 64 hours. The solid line is the fit to the Gaussian distributions.}
    \label{SMphase}
\end{figure}

The method to calibrate $\phi$ is the same as Ref.\cite{liang2022experimental}. To calibrate the phase $\phi$, one kind of  $b_{\mathrm{other}}$ is temporarily introduced.  A copper wire is glued to the bender. A direct current is applied to the copper wire, and a magnetic field that is in phase with the position of sphere $r_M(t)$ is generated due to the vibration. To ensure that exotic velocity-dependent interaction does not influence the calibration, the amplitude of the magnetic field from the copper wire is 3 nT, which is much larger than the magnetic field when the current is not applied. As shown in Fig.~\ref{SMphase}(a), before adjusting the phase $\phi$, the outputs of both channels are non-zero when a direct current is applied. After calibration, the phase of the reference signal in the second lock-in amplifier is tuned to $51^\circ$ where the quadrature component is zeros when the direct current is applied. The result is shown in Fig.\ref{SMphase}(b), and the uncertainty of the phase $\phi$ is $|\Delta\phi|\leq4^\circ$. For $|\Delta\phi|=4^\circ$, the output of the quadrature channel contains 99.8\% of the velocity-dependent $b_{\mathrm{eff}}$ and 7\% of the position-dependent $b_{\mathrm{other}}$.

The current is turned off when performing the search for the exotic interaction. The output from the in-phase channel and the quadrature channel are shown in Fig.~\ref{SMphase}(c)(d), which both obey the Gaussian distribution. The fundamental frequency component of $b_{\mathrm{other}}$ is $a_1^{exp}=2.3\pm3.2$ pT and the fundamental frequency component of $b_{\mathrm{eff}}$ is $b_1^{exp}=1.0\pm3.2$ pT. The result that the mean value of the measured magnetic field is less than the standard error of the mean shows that there is no significant magnetic field in phase with the velocity or the position.

\section{Statistical properties of the results}
\label{SM4}
The correlation coefficient $\rho(A,B)$ characterizes the correlation between ${A_i}$ and ${B_i}$. The definition of the correlation coefficient is:
\begin{equation}
\begin{aligned}
\rho(A,B)=\frac{1}{N-1}\sum_{i=1}^{N}(\frac{A_i-\mu_A}{\sigma_A})(\frac{B_i-\mu_B}{\sigma_B})
\end{aligned}
\end{equation}
where $\mu_A$($\mu_B$) and $\sigma_A$($\sigma_B$) are the mean and standard deviation of A (B).  The correlation coefficient of the in-phase channel  and the quadrature channel is $\rho(I,Q)=0.0041\pm0.0044$.  Therefore, the output from two channels is considered to be uncorrelated.

The normalized autocorrelation of A is:
\begin{equation}
\begin{aligned}
R_{A}(m)=\frac{1}{\sum_{n=0}^{N-1}A_{n}A_n^*}\times\begin{cases}
\sum_{n=0}^{N-m-1}A_{n+m}A_n^* &,m\ge0\\
R_{A}^*(-m)&,m\le0
\end{cases}
\end{aligned}
\end{equation}
$|R_{I}(m)|<=0.0094$ and  $|R_{Q}(m)|<=0.0098$ when $m\neq 0$. There is no correlation between the points. Therefore, the standard error of the mean can be calculated by $\sigma/\sqrt{N}$.

The statistical properties of the output from two channels are listed in Table\ref{static}.

\begin{table}[h!]
 \caption{Statistical properties of the results}
 \label{static}
\begin{tabular}{l |c| c}
\hline
\hline
Channel&I&Q\\
\hline
Number&203495&203495\\
Mean (pT)&2.3 &1.0\\
Standard error (nT)&1.4&1.4\\
Standard error of mean (pT)&3.2&3.2\\
\hline
\hline
\end{tabular}
\end{table}

\section{Systematicerrors}
 \label{SM5}
\subsection{Uncertainty in the angle $\alpha$ between the velocity and the $\hat{y}$ direction}
The angle $\alpha$ between the velocity and the $\hat{y}$ direction is $0\pm3^\circ$, which is determined by the edge of the bender and the edge of the diamond. The correction to the coupling parameter $f_\perp$ is from $0.0\times 10^{-14}$ to $0.2\times 10^{-14}$ at $\lambda=100\ \upmu$m.

\subsection{Uncertainty in the diameter}
The diameter of the lead sphere is measured to be $897\pm 3\ \upmu$m with  a micrometer. The correction to the coupling parameter $f_\perp$ is $\pm0.5\times 10^{-14}$ at $\lambda=100\ \upmu$m.

\subsection{Uncertainty in the distance}
 The distance between the bottom of the sphere and the surface of the diamond is controlled by a vertically installed piezo motor carrying the piezoelectric bender. The feedback from the piezoelectric bender is monitored when the sphere slowly moves down with a tiny vibration amplitude. The position where the sphere approaches the surface of the diamond is determined where the feedback suddenly varies\cite{liang2022experimental}. The distance between the bottom of the sphere and the surface of the diamond is set to $5.0\pm0.5\ \upmu$m. The uncertainty is determined by measuring the position where the sphere approaches the surface of the diamond multiple times. The correction to the coupling parameter $f_\perp$ is $\pm0.8\times10^{-14}$ at $\lambda=100\ \upmu$m.

\subsection{Uncertainty in the thickness}
The thickness of the NV layer is determined by the thickness difference before and after diamond growth. The NV layer is fabricated on a diamond whose thickness is $551\pm1\ \upmu$m. The thickness of the diamond after the growth of the NV layer is $574 \pm1\ \upmu$m which indicates the thickness of the NV layer is $23\pm1\ \upmu$m. The correction to the coupling parameter $f_\perp$ is $\pm0.8\times10^{-14}$ at $\lambda=100\ \upmu$m.

\subsection{Uncertainty in the amplitude}
The vibration amplitude is $538\pm3$ nm which is measured by a commercial laser vibrometer (Sunny Optical, LV-S01). The correction to the coupling parameter $f_\perp$ is $\pm0.9\times10^{-14}$ at $\lambda=100\ \upmu$m.

\subsection{Uncertainty in the deviation}
The deviation in X-Y plane is $0\pm10\ \upmu$m according to the CCD images. The correction to the coupling parameter $f_\perp$ is from $-0.8\times10^{-14}$ to $0.9\times 10^{-14}$ at $\lambda=100\ \upmu$m.
\subsection{Uncertainty in the nucleon density}

The density the lead sphere is measured to be $\rho= 11.381\pm0.036\ \mathrm{g/cm^3}$. The corresponding nucleon density is $6.82\pm0.02\ \times10^{30}\ \mathrm{m^{-3}}$ and the correction to the coupling parameter $f_\perp$ is  $\pm0.5\times10^{-14}$ at $\lambda=100\ \upmu$m.

\subsection{Uncertainty in the coefficient  $\eta$}
The coefficient $\eta$  between the magnetometer output voltage signal and the magnetic field is measured to be $\eta=47.2\pm0.2\ \upmu$T/V. The correction to the coupling parameter $f_\perp$ is $\pm0.7\times10^{-14}$ at $\lambda=100\ \upmu$m.

The systematic error is much less than the statistical error. With the fiducial probability of 95\%,  the bound of $f_\perp$ with $\lambda=100\ \upmu$m is $\lvert f_\perp \rvert \leq 1.1\times 10^{-11}$ when both statistical and systematic errors are taken into account.

 \section{Comparison with the non-zero result observed in the single-NV-based experiment}
 \label{SM6}
 \begin{figure}[htbp]
\centering
\includegraphics[width=0.9\columnwidth]{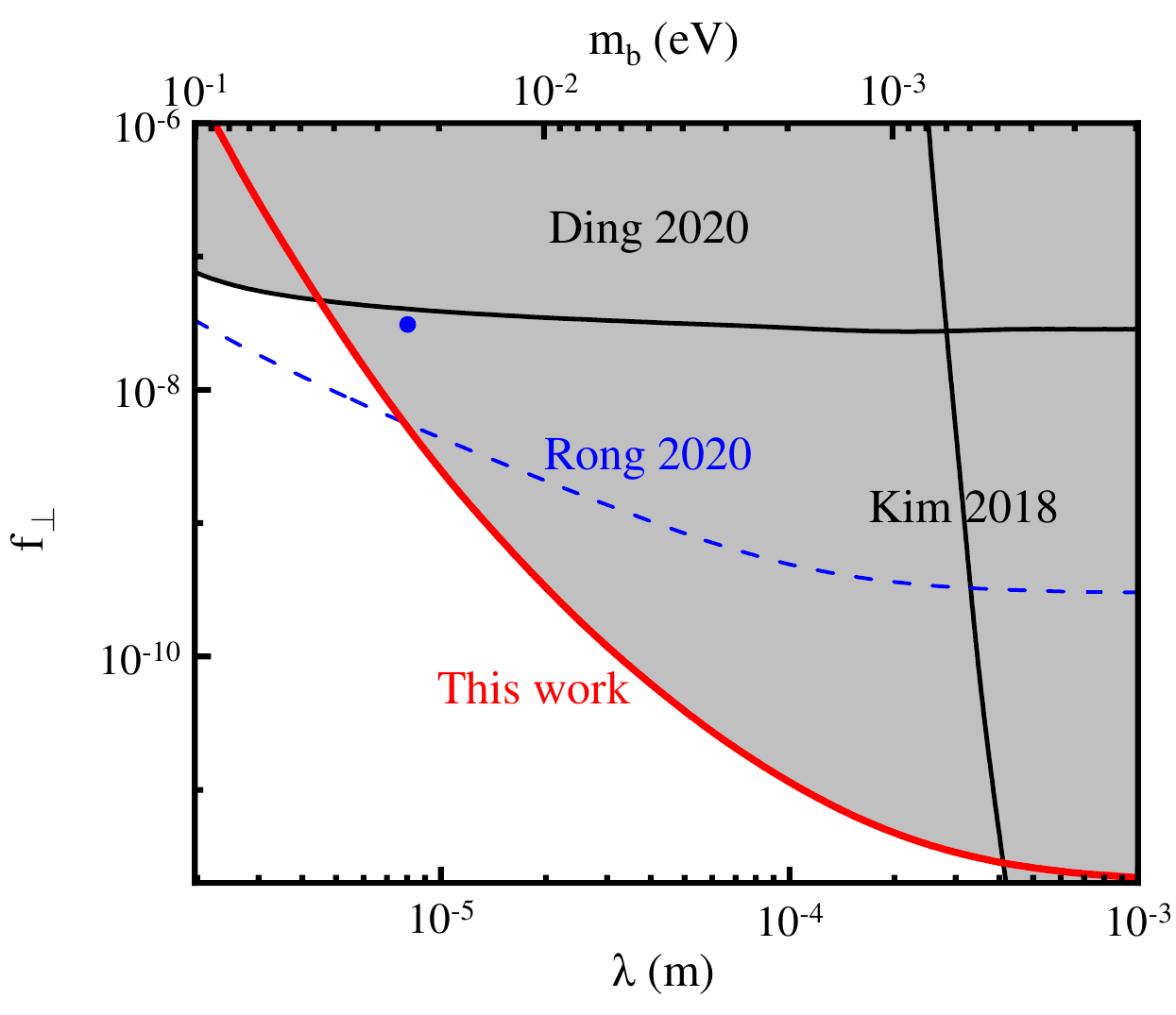}
    \caption{The upper bounds of the coupling parameter $f_\perp$ as a function of the force range $\lambda$ and the boson mass $m_b$. Black lines are upper bounds obtained from previous experiments \cite{kim2018experimental,ding2020constraints}. The red line is the upper limit established by our experiment. The blue dash line is the experimental precision in Ref.\cite{rong2020observation}, while the blue dot denotes the non-zero result in Ref.\cite{rong2020observation}. The gray-filled area shows the excluded values.}
    \label{Bounds}
\end{figure}

Recently, a non-zero result was reported in an experimental searching for exotic spin-denpendent interaction \cite{rong2020observation}. The upper bounds established by this work and previous experiments are shown in Fig.\ref{Bounds}.  The suggested coupling parameter $f_\perp=3.93\times10^{-8}$ with the force range $\lambda=8.07\ \upmu$m in Ref. \cite{rong2020observation} is shown as a blue dot. Our experiment sets bounds more stringent than the experimental precision in Ref.\cite{rong2020observation} in force range larger than 8 micrometers. Our results partially exclude the non-zero result and suggest that the non-zero result in Ref. \cite{rong2020observation} may come from instrumental artifacts.

\end{document}